
\def\t{\textstyle}


\baselineskip 14pt
\vglue .5cm
\centerline
{\bf A SYSTEM WITH A RECURSION OPERATOR,}
\centerline{\bf BUT ONE HIGHER LOCAL SYMMETRY}
\vskip .5cm
\centerline{Ay\c se H\"umeyra Bilge}
\vskip .2cm
\centerline{Department of Mathematics, Anadolu University}
\centerline{26470 Eskisehir, Turkey}
\centerline{Present Address: Department of Mathematics, Istanbul Technical
University}
\centerline{ Maslak, Istanbul, TURKEY}
\vskip .5cm
\centerline{\bf ABSTRACT}
\vskip .2cm

We construct a recursion operator for the system $(u_t,v_t)=(u_4+v^2,{1\over
5}v_4)$, for which only one local
symmetry is known and we show that the action of the recursion operator on
$(u_t,v_t)$ is a local function.

\vskip 1cm

\baselineskip=18 pt
\noindent
{\bf 1. Introduction. }
\vskip .2cm
We consider the system of evolution equations
$$\eqalign{u_t&=u_m+v^2\cr
           v_t&=\lambda v_m, \quad \lambda \ne 0.\cr}\eqno(1.1)
$$
These systems are studied in [1] as
examples of systems for which only one local symmetry is known.
It is shown that
for arbitrary $\lambda$, they have classical and Lie-point symmetries
$(\sigma,\eta):$
$$(1,0),\quad\quad(2u,v),\quad\quad(u_1,v_1),\quad\quad(u_t,v_t),\eqno(1.2)$$
and it is shown that higher symmetries are of the form
$$(a u_n+\t\sum_{i=0}^s \beta_iv_iv_{n-m-i}, bv_n),\quad
s=[(n-m)/2].\eqno(1.3)$$
It is also shown that for $m=4$,
and $\lambda={1\over 5}$ the system admits the   6th order symmetry
$$(\t{11\over 25}u_6 +v_2v+\t{4\over 5} v_1^2, \t{1\over 25}v_6)\eqno(1.4)$$
but no higher local symmetry of this equation could have been found
up to order 53 [1].

We will show that the system (1.1) admits always a ``formal" recursion
operator
$R$, in the sense of being the solution of the operator equation
$$R_t+[R,F']=0\eqno(1.5)$$
where $F'$ is the Frechet derivative of $F=\pmatrix{u_m+v^2\cr \lambda v_m}$.
Then, for  $m=4$ and $Ord(R)=2$, we determine the coefficients of
$R$ explicitly and
by appropriate choice of free parameters we show that $R$ acting on
the symmetry $(u_t,v_t)$
gives the 6th order symmetry (1.4).

\vskip .5cm
\noindent
{\bf 2. Computation of the recursion operator.}
\vskip .2cm

It can be seen that the symmetries $(\sigma,\eta)$ satisfy the equation
$$\eqalign{
\sigma_t&=D^m\sigma+2v\eta,\cr
\eta_t  &=\lambda D^m\eta.\cr}\eqno(2.1)$$
The recursion operator $R$ is a matrix operator of the form
$$R=\pmatrix{A&B\cr  C&D\cr}.    $$
The operator equation (1.5) gives
$$\eqalignno{&A_t+AD^m-D^mA=0,&(2.2a)\cr
           &B_t+2Av+\lambda B D^m-D^mB-2vE=0,&(2.2b)\cr
           &C_t+CD^m-\lambda D^mC=0,&(2.2c)\cr
           &E_t+2Cv+ \lambda ED^m-\lambda D^mE=0.&(2.2d)\cr}$$
We first show that $C=0$.
If $Ord(C)=k$, then the top term of (2.2c) is of
order $k+m$, with coefficient $(\lambda-1)C_k$, which implies that $C=0$. Then
it follows that $A$ and $E$ can be arbitrary constant coefficient operators
and the problem is reduced to solving
$$B_t+\lambda BD^m-D^m B +2Av-2vE=0.\eqno(2.3)$$
If $Ord(B)=n$, $Ord(A)=k$ and $Ord(E)=l$, then it can be seen that provided
that $n+m=k$ or $n+m=l$, the operator equation in (2.3)  can always be solved
for the coefficients of  $B$ recursively, because the top term
involves the top term of $B$
algebraically, with a nonzero coefficient.
This situation is in constrast with
the scalar case where the top term of the operator equation always involves the
first derivative of the top  coefficient to be determined at that stage, and
the solvability of condition gives  conserved densities that
lead to a  classification.

Thus the system (1.1) possesses a recursion operator for any $m$ and any
$\lambda
\ne 1$. We recall the existence of a ``formal symmetry" is proposed as an
integrability test in [2].   A formal symmetry is a pseudo-differential
operator that satisfies the equation (1.5) up to a certain order, hence the
existence of a  recursion operator implies integrability in this sense. We note
that in [2], the integrability test for systems of evolution equations
involves a formal diagonalization procedure that first transforms the
differential function $F$ to a pseudo-differential operator, then the problem
is reduced to finding formal symmetries for a set of scalar equations.
Here we have used a direct
computation of the recursion operator, because we were interested not only in
proving the existence of a formal symmetry, but  in finding the recursion
operator explicitly.

In the following we will assume that
$$A=aD^2,\quad\quad E=bD^2.\eqno(2.4)$$
We note that the expansions of a pseudo-differential operator of the
forms $B=\sum B_i D^{-i}$  or $B=\sum D^{-i} B_i$ are equivalent. Here we
choose the second formulation. It can easily
be seen that $B$ is of the form
$$B=\sum_{j=0}^\infty b_j D^{-j-2}v_j\eqno(2.5)$$
where the $b_j$'s are constants.

Substituting $B$ in (2.3) and using the  commutation relation
$$D^{-1}\varphi=\varphi D^{-1}-D^{-1}\varphi_1 D^{-1}\eqno(2.6)$$
one can compute the coefficients of $D^2$, $D$, $1$, $D^{-1}$, $\dots$, and
obtain the following equations for the $b_i$'s.

$$\eqalign{&(\lambda-1)b_0+2a-2b=0,\cr
           &(\lambda-1)b_1-2(\lambda+1)b_0+4a=0,\cr
           &(\lambda-1)b_2-(3\lambda+1)b_1++(3\lambda-1)b_0+2a=0,\cr
           &(\lambda-1)b_3-4\lambda b_0+6\lambda b_1-4\lambda b_2=0,\cr
           }\eqno(2.7)$$
and the recurrence relation
$$(\lambda-1)b_n-4\lambda b_{n-1}+6\lambda b_{n-2}-4\lambda b_{n-3}
               +2\lambda b_{n-4}=0.\eqno(2.8)$$
From this recurrence relation, it is easy to see that the series cannot
terminate.

\vskip .5cm
\noindent
{\bf 3. Existence of one local symmetry.}

We will show that $B$ acting on $v_4$ results in a local function. For this we
will need the following.

\proclaim Lemma. Let
$S=\sum_{n=4}^\infty b_nD^{-n-2}v_nv_4.          $
Assume that
$$b_n=a_1 b_{n-1}+a_2b_{n-2}+a_3b_{n-3}+a_4b_{n-4} \quad \quad n\ge
8\eqno(3.1)$$
with
$a_1+a_2+a_3+a_4=0,$
and
$$(2+a_4)b_4+(1+a_3+a_4)b_5+(1+a_2+a_3+a_4)b_6+b_7=0.\eqno(3.2)$$
If the condition (3.2) is invariant under the transformation
$$\pmatrix{b_4\cr b_5 \cr b_6\cr b_7\cr}\to
\pmatrix{a_4 & a_3+a_4 & 1+a_2+a_3+a_4 & 1  \cr
         a_4 & a_3+a_4 &   a_2+a_3+a_4 & 1  \cr
         a_4 & a_3+a_4 &   a_2+a_3+a_4 & 0  \cr
           0 &     a_4 &       a_3+a_4 & a_2+a_3+a_4 \cr}
  \pmatrix{b_4\cr b_5 \cr b_6\cr b_7\cr}\eqno(3.3)$$
  then $S$ is zero.

\noindent
{\it Proof.}  Recall that $D^{-1} v_nv_k=v_{n-1}v_k-D^{-1}v_{n-1}v_{k+1}$.
Hence by successive iterations of this formula, $D^{-n-2}v_nv_4$ will be a
linear combination of
$\{D^{-2n+4}v_n^2\}_{n=4}^{\infty}$ $=\{D^{-6}v_4^2,D^{-8}v_5^2,\dots\}$. We
will show that the coefficient of $D^{-2n+4}v_n^2$ vanishes for all $n\ge
4$. We rewrite $S$ by substituting the recurrence relations for $n\ge 8$.
$$
S=\sum_{n=4}^7 b_nD^{-n-2}v_nv_4
    +\sum_{n=8}^\infty \sum_{k=1}^4 a_k b_{n-k}D^{-n-2}v_nv_4.$$
    Note that the series $S_4=\sum_{n=8}^\infty b_{n-4}D^{-n-2}v_nv_4$ can be
integrated to give
$$\eqalignno{
  S_4&=\sum_{n=8}^\infty b_{n-4}D^{-n-1}v_{n-1}v_4
     -\sum_{n=8}^\infty b_{n-4}D^{-n-2}v_{n-1}v_5\cr
     &=b_4D^{-9}v_7v_4+\sum_{n=9}^\infty b_{n-4}D^{-n-1}v_{n-1}v_4
     -\sum_{n=8}^\infty b_{n-4}D^{-n-2}v_{n-1}v_5\cr
     &=b_4D^{-9}v_7v_4+\sum_{n=8}^\infty b_{n-3}D^{-n-2}v_{n  }v_4
     -\sum_{n=8}^\infty b_{n-4}D^{-n-2}v_{n-1}v_5.\cr}$$
     By combining the first series above with the series $a_3\sum_{n=8}^\infty
b_{n-3}D^{-n-2}v_nv^4$, we obtain
$$\eqalignno{
S=&\sum_{n=4}^7 b_nD^{-n-2}v_nv_4+a_4b_4D^{-9}v_7v_4
     -a_4\sum_{n=8}^\infty b_{n-4}D^{-n-2}v_{n-1}v_5\cr
   & +a_1\sum_{n=8}^\infty b_{n-1}D^{-n-2}v_{n}v_4
     +a_2\sum_{n=8}^\infty b_{n-2}D^{-n-2}v_{n}v_4\cr
   &  +(a_3+a_4)\sum_{n=8}^\infty b_{n-3}D^{-n-2}v_{n}v_4\cr}$$
Repeating the same procedure we obtain
$$\eqalignno{
S=&\sum_{n=4}^7
b_nD^{-n-2}v_nv_4+a_4b_4D^{-9}v_7v_4+(a_3+a_4)b_5D^{-9}v_7v_4\cr
  &  -a_4\sum_{n=8}^\infty b_{n-4}D^{-n-2}v_{n-1}v_5
    -(a_3+a_4)\sum_{n=8}^\infty b_{n-3}D^{-n-2}v_{n-1}v_5\cr
  &  +a_1\sum_{n=8}^\infty b_{n-1}D^{-n-2}v_{n}v_4
    +(a_2+a_3+a_4)\sum_{n=8}^\infty b_{n-2}D^{-n-2}v_{n}v_4\cr}$$
  and finally
  $$\eqalignno{
S=&\sum_{n=4}^7
b_nD^{-n-2}v_nv_4+a_4b_4D^{-9}v_7v_4
+(a_3+a_4)b_5D^{-9}v_7v_4\cr
&+(a_2+a_3+a_4)b_6D^{-9}v_7v_4
 -a_4\sum_{n=8}^\infty b_{n-4}D^{-n-2}v_{n-1}v_5 \cr
&    -(a_3+a_4)\sum_{n=8}^\infty b_{n-3}D^{-n-2}v_{n-1}v_5
    -(a_2+a_3+a_4)\sum_{n=8}^\infty b_{n-2}D^{-n-2}v_{n-1}v_5\cr
&    +(a_1+a_2+a_3+a_4)\sum_{n=8}^\infty b_{n-1}D^{-n-2}v_{n}v_4\cr}$$
 But the last series vanishes because the sum of the $a_i$'s is zero. By
rearranging we obtain
$$\eqalignno{
S=&b_4D^{-6}v_4^2+b_5D^{-7}v_5v_4+b_6D^{-8}v_6v_4\cr
  &+[b_7+a_4b_4+(a_3+a_4)b_5+(a_2+a_3+a_4)b_6]D^{-9}v_7v_4\cr
  &-\sum_{n=8}^\infty[a_4b_{n-4}+(a_3+a_4)b_{n-3}+(a_2+a_3+a_4)b_{n-2}]
  D^{-n-2}v_{n-1}v_5\cr}$$
Using
$$\eqalignno{
D^{-7}v_5v_4&=\t{1\over 2}D^{-6}v_4^2,\cr
D^{-8}v_6v_4&=\t{1\over 2}D^{-6}v_4^2-D^{-8}v_5^2,\cr
D^{-9}v_7v_4&=\t{1\over 2}D^{-6}v_4^2-D^{-8}v_5^2-D^{-9}v_6v_5,\cr}$$
we get
$$\eqalignno{
S=&\big[b_4+\t{1\over 2}b_5+\t{1\over 2}b_6+\t{1\over
2}\big(b_7+a_4b_4+(a_3+a_4)b_5+(a_2+a_3+a_4)b_6\big)\big]D^{-6}v_4^2\cr
&-[b_6+b_7+a_4b_4+(a_3+a_4)b_5+(a_2+a_3+a_4)b_6]D^{-8}v_5^2\cr
&-[b_7+a_4b_4+(a_3+a_4)b_5+(a_2+a_3+a_4)b_6]D^{-9}v_6v_5\cr
&-\sum_{n=8}^\infty[a_4b_{n-4}+(a_3+a_4)b_{n-3}+(a_2+a_3+a_4)b_{n-2}]
  D^{-n-2}v_{n-1}v_5\cr}$$
The coefficient of $D^{-6}v_4^2$ is just the condition in (3.2), hence this
term vanishes by assumption. Thus $S$ is now of the form
$$S=-\sum_{n=4}^\infty c_n D^{-n-4}v_{n+1}v_5$$
where
$$\eqalign{
c_4&=b_6+b_7+a_4b_4+(a_3+a_4)b_5+(a_2+a_3+a_4)b_6,\cr
c_4&=    b_7+a_4b_4+(a_3+a_4)b_5+(a_2+a_3+a_4)b_6,\cr}$$
and
$$
c_n=        a_4b_{n-2}+(a_3+a_4)b_{n-1}+(a_2+a_3+a_4)b_n, \quad n\ge 6.$$
It is easy to see that the $c_n$'s satisfy the same recursion relation as the
$b_n$'s, and they are related to them by the transformation formula (3.3).
Hence the same procedure can be repeated to show that the coefficient of
$D^{-8}v_5^2$ is zero. As all the arguments can be repated by replacing $v_4$
with $v_k$ it can be shown by induction that the coefficient of
$D^{-2(n-1)}v_n^2$ in $S$ is zero.
\vskip .2cm

We  will now show that the conditions of the Lemma hold for our system.
In our case
$$a_1=-{4\lambda\over 1-\lambda},\quad
  a_2= {6\lambda\over 1-\lambda},\quad
  a_3=-{4\lambda\over 1-\lambda},\quad
  a_4= {2\lambda\over 1-\lambda},\quad\eqno(3.4)$$
Let $A$ denote the matrix in (3.3). It can be seen that the minimal polynomial
of $A$ is
$$A^2+{4\lambda\over 1-\lambda}A+{2\lambda\over 1-\lambda}=0.\eqno(3.5)$$
The condition (3.2) can be interpreted as the orthogonality of the vector
$b=(b_4,b_5,b_6,b_7)$ with the  fixed vector
$d=(2+a_4,1+a_3+a_4,1+a_2+a_3+a_4,1),$. If $b$ is chosen such that
$d^tb=0$ and $d^tAb=0$, then as the minimal polynomial of $A$ has order 2, the
condition  $d^tA^nb=0$ will hold for all $n$. This means that the initial
condition (3.2) will hold for all iterations.

The conditions $d^tb=0$ and $d^tAb=0$ determine $b_7$ and $b_6$ as
$$\eqalign{
b_7=&{2\over (\lambda-1)^2}[2(1+5\lambda)b_4+(1+7\lambda)b_5]\cr
b_6=&{1\over  \lambda-1   }[6            b_4+(3+ \lambda)b_5]\cr}\eqno(3.6)$$
On the other hand $b_6$ and $b_7$ are already given by recurrence relations.
The compatibility of these expressions give a homogeneous system for $a$ and
$b$. The determinant of this system is zero only for $\lambda={1\over 5}$. In
this case $a$ is determined as ${11\over 5}b$. Substituting these values in
(2.7), $b_i$ $i=0,\dots 3$ can be computed, and it can be seen that
$Bv_4=b(3vv_2-2 v_1^2)$. In particular, the coefficient of
$D^{-4}v_2^2$ vanishes without giving any further restriction.
Then, the symmetry $R(u_t,v_t)=(aD^2(u_4+v^2)+B({1\over 5}v_4), bD^2({1\over
5}v_4)$ can ve computed, and is equal to the expression in (1.4).

We present the first few $b_i$'s
for $b={1\over 5}$ below.
$$\matrix{
b_0=0.6&
b_1=0.4&
b_2=0 &
b_3=0 &
b_4= - 0.1\cr
b_5=0.3   &
b_6= - 0.45 &
b_7=1       &
b_8= - 2.025 &
b_9=4.125 \cr
b_10= - 8.388       &
b_11=17.1           &
b_12= - 34.82       &
b_13=70.92          &
b_14= - 144.4       \cr
b_15=294.2          &
b_16= - 599.2       &
b_17=1220.0         &
b_18= - 2486.0      &
b_19=5062.0         \cr}$$

\vskip.5cm
\noindent
{\bf 4. Conclusion.}

We have shown that the system (1.1) has always a recursion operator which
is an infinite series in $D^{-i}$ and we have calculated the coefficients for
the case $m=4$ given in (2.7) and (2.8). From the recurrence relation (2.8)
it can be seen that the series cannot terminate, and the atempts to write the
recursion operator in closed form were not succsessful. Neverthless the series
$Bv_4$ teminates and furthermore it is a local function. The action of
$B$ on higher symmetries in general do not give a series that terminates. The
condition for the series to terminate is the orthogonality of the vectors
$(b_n,b_{n+1},b_{n+2},b_{n+3})$ with the vectors $d=({5\over 4},{1\over
4},1,{1\over 2})$ and $dA^t=({5\over 2},-2,7,4)$. These conditions do not
follow from the recursion relations, and they are not satisfied for the
next few symmetries.

 The existence of a recursion
operator, a formal symmetry or an eigenvalue problem are more or less
equivalent methods that lead to a classification [2,3,4]. This example shows
that the the existence of an infinite number of local symmetries is more
restrictive.

\vskip .5cm
\noindent
{\bf Acknowledgements.}
The author would like to thank Professor P.J. Olver  for suggesting this
problem. This work is partially supproted by the study group \c CG-1 of the
Scientific and Technical Research Council of Turkey.

\vskip .5cm
Note Added in Submission to solv-int: This problem has also been studied
by F. Beukers, J.A. Sanders and J.P. Wang in ``One symmetry does not imply
integrability", {\it Journal of Differential Equations}, {\bf 146}, pp.
251-260, (1998).

\vskip .5cm
 \noindent
{\bf References.}  \vskip .2cm

\baselineskip 14pt
\noindent
 1. I.M.Bakirov, {\it On the symmetries of some system of} {\it evolution
equations}, Pre\-print.
\vskip .1cm

\noindent
2. A.V.Mikhailov, A.B.Shabat and V.V.Sokolov, {\it The symmetry approach to the
classification} {\it of integrable equations}, in ``What is Integrability?",
Ed. V. E. Zakharov, Springer, Berlin, 1990.
 \vskip .1cm

\noindent
3.M.G\"urses, A.Karasu and A.Sat\i r, {\it Linearization as a new test for
integrability}, in ``Nonlinear Evolution Equations and Dynamical Systems",
(Proc. NEEDS '91, Gallipoli, June 1991), Eds.
M.Boiti, L.Martina and F. Pampinelli, World Scientific, Singapore, 1991.
\vskip .1cm

\noindent
4. A.H.Bilge, {\it On the equivalence of linearization and formal symmetries as
integrability test for evolution equations}, J. Phys. A: Math. Gen. {\bf 26},
7511-7519, 1993.

           \end